\documentstyle[twoside,fleqn,espcrc2]{article}

\newcommand{\AmS}{{\protect\the\textfont2
  A\kern-.1667em\lower.5ex\hbox{M}\kern-.125emS}}

\newcommand{\pr}{{Phys.\ Rev.\/}}

\newcommand{\np}{{Nucl.\ Phys.\/}}
\newcommand{\pl}{{Phys.\ Lett.\/}}

\newcommand{\beq}{\begin{equation}}
\newcommand{\eeq}{\end{equation}}
\newcommand{\backn}{\begin{ack}}
\newcommand{\eackn}{\end{ack}}
\title{Field Strength and Monopoles in Dual $U(1)$ Lattice Gauge Theory
\thanks{Supported in part by Fonds zur F\"orderung der wissenschaftlichen Forschung, Proj.~P11156-PHY}}
 
\author{Martin Zach, Manfried Faber and Peter Skala  \\
Institut f\"ur Kernphysik, Technische Universit\"at Wien, A-1040 Vienna, Austria}
       
\begin{document}

\begin{abstract}
In any Abelian gauge theory with an action periodic in the link variables
one can perform a duality transformation not only in the partition function,
but also in correlation functions including Polyakov loops.
The calculation of expectation values in the confinement phase, like electric field strength or monopole currents in the presence of external charges,
 becomes significantly more efficient simulating the dual theory. We demonstrate this using the ordinary Wilson action.
This approach also allows a quantitative analysis of the dual superconductor model, because the dual transformed $U(1)$ theory can be regarded as limit of a dual non-compact Abelian Higgs model. In this way we also try to interpret the behaviour of monopole condensate and string fluctuations. Finally we present some applications for simulating the dual $U(1)$ gauge theory.
\end{abstract}

\maketitle
\begin{figure*}[t]
\vspace{-0.1cm}
\centerline{\input{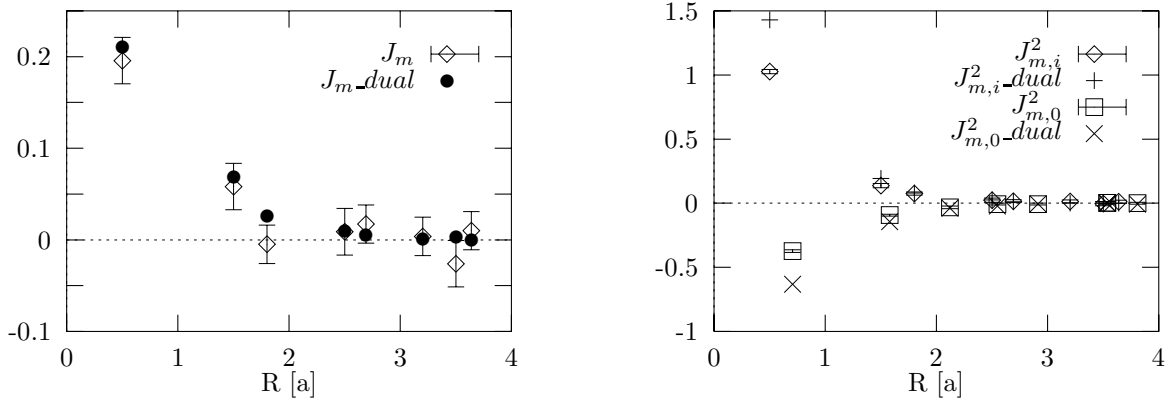}
\setlength{\unitlength}{0.1bp}
\special{!
/gnudict 40 dict def
gnudict begin
/Color false def
/Solid false def
/gnulinewidth 2.000 def
/vshift -33 def
/dl {10 mul} def
/hpt 31.5 def
/vpt 31.5 def
/M {moveto} bind def
/L {lineto} bind def
/R {rmoveto} bind def
/V {rlineto} bind def
/vpt2 vpt 2 mul def
/hpt2 hpt 2 mul def
/Lshow { currentpoint stroke M
  0 vshift R show } def
/Rshow { currentpoint stroke M
  dup stringwidth pop neg vshift R show } def
/Cshow { currentpoint stroke M
  dup stringwidth pop -2 div vshift R show } def
/DL { Color {setrgbcolor Solid {pop []} if 0 setdash }
 {pop pop pop Solid {pop []} if 0 setdash} ifelse } def
/BL { stroke gnulinewidth 2 mul setlinewidth } def
/AL { stroke gnulinewidth 2 div setlinewidth } def
/PL { stroke gnulinewidth setlinewidth } def
/LTb { BL [] 0 0 0 DL } def
/LTa { AL [1 dl 2 dl] 0 setdash 0 0 0 setrgbcolor } def
/LT0 { PL [] 0 1 0 DL } def
/LT1 { PL [4 dl 2 dl] 0 0 1 DL } def
/LT2 { PL [2 dl 3 dl] 1 0 0 DL } def
/LT3 { PL [1 dl 1.5 dl] 1 0 1 DL } def
/LT4 { PL [5 dl 2 dl 1 dl 2 dl] 0 1 1 DL } def
/LT5 { PL [4 dl 3 dl 1 dl 3 dl] 1 1 0 DL } def
/LT6 { PL [2 dl 2 dl 2 dl 4 dl] 0 0 0 DL } def
/LT7 { PL [2 dl 2 dl 2 dl 2 dl 2 dl 4 dl] 1 0.3 0 DL } def
/LT8 { PL [2 dl 2 dl 2 dl 2 dl 2 dl 2 dl 2 dl 4 dl] 0.5 0.5 0.5 DL } def
/P { stroke [] 0 setdash
  currentlinewidth 2 div sub M
  0 currentlinewidth V stroke } def
/D { stroke [] 0 setdash 2 copy vpt add M
  hpt neg vpt neg V hpt vpt neg V
  hpt vpt V hpt neg vpt V closepath stroke
  P } def
/A { stroke [] 0 setdash vpt sub M 0 vpt2 V
  currentpoint stroke M
  hpt neg vpt neg R hpt2 0 V stroke
  } def
/B { stroke [] 0 setdash 2 copy exch hpt sub exch vpt add M
  0 vpt2 neg V hpt2 0 V 0 vpt2 V
  hpt2 neg 0 V closepath stroke
  P } def
/C { stroke [] 0 setdash exch hpt sub exch vpt add M
  hpt2 vpt2 neg V currentpoint stroke M
  hpt2 neg 0 R hpt2 vpt2 V stroke } def
/T { stroke [] 0 setdash 2 copy vpt 1.12 mul add M
  hpt neg vpt -1.62 mul V
  hpt 2 mul 0 V
  hpt neg vpt 1.62 mul V closepath stroke
  P  } def
/S { 2 copy A C} def
end
}
\begin{picture}(2339,1511)(0,0)
\special{"
gnudict begin
gsave
50 50 translate
0.100 0.100 scale
0 setgray
/Helvetica findfont 100 scalefont setfont
newpath
-500.000000 -500.000000 translate
LTa
480 735 M
1676 0 V
480 251 M
0 1209 V
LTb
480 251 M
63 0 V
1613 0 R
-63 0 V
480 493 M
63 0 V
1613 0 R
-63 0 V
480 735 M
63 0 V
1613 0 R
-63 0 V
480 976 M
63 0 V
1613 0 R
-63 0 V
480 1218 M
63 0 V
1613 0 R
-63 0 V
480 1460 M
63 0 V
1613 0 R
-63 0 V
480 251 M
0 63 V
0 1146 R
0 -63 V
899 251 M
0 63 V
0 1146 R
0 -63 V
1318 251 M
0 63 V
0 1146 R
0 -63 V
1737 251 M
0 63 V
0 1146 R
0 -63 V
2156 251 M
0 63 V
0 1146 R
0 -63 V
480 251 M
1676 0 V
0 1209 V
-1676 0 V
480 251 L
LT0
1913 1297 D
690 1231 D
1109 801 D
1235 771 D
1528 746 D
1608 742 D
1822 741 D
1947 733 D
2005 740 D
1853 1297 M
180 0 V
-180 31 R
0 -62 V
180 62 R
0 -62 V
690 1223 M
0 16 V
-31 -16 R
62 0 V
-62 16 R
62 0 V
1109 793 M
0 16 V
-31 -16 R
62 0 V
-62 16 R
62 0 V
95 -44 R
0 12 V
-31 -12 R
62 0 V
-62 12 R
62 0 V
262 -39 R
0 16 V
-31 -16 R
62 0 V
-62 16 R
62 0 V
49 -17 R
0 11 V
-31 -11 R
62 0 V
-62 11 R
62 0 V
183 -12 R
0 11 V
-31 -11 R
62 0 V
-62 11 R
62 0 V
94 -22 R
0 16 V
-31 -16 R
62 0 V
-62 16 R
62 0 V
27 -7 R
0 12 V
-31 -12 R
62 0 V
-62 12 R
62 0 V
LT1
1913 1197 A
690 1426 A
1109 828 A
1235 773 A
1528 747 A
1608 740 A
1822 735 A
1947 735 A
2005 735 A
LT0
1913 1097 B
776 554 B
1142 690 B
1369 719 B
1548 731 B
1701 731 B
1962 736 B
1962 736 B
2076 735 B
1853 1097 M
180 0 V
-180 31 R
0 -62 V
180 62 R
0 -62 V
776 549 M
0 10 V
745 549 M
62 0 V
-62 10 R
62 0 V
335 127 R
0 7 V
-31 -7 R
62 0 V
-62 7 R
62 0 V
196 21 R
0 10 V
-31 -10 R
62 0 V
-62 10 R
62 0 V
148 4 R
0 7 V
-31 -7 R
62 0 V
-62 7 R
62 0 V
122 -7 R
0 7 V
-31 -7 R
62 0 V
-62 7 R
62 0 V
230 -3 R
0 9 V
-31 -9 R
62 0 V
-62 9 R
62 0 V
-31 -9 R
0 7 V
-31 -7 R
62 0 V
-62 7 R
62 0 V
83 -8 R
0 7 V
-31 -7 R
62 0 V
-62 7 R
62 0 V
LT3
1913 997 C
776 429 C
1142 665 C
1369 716 C
1548 724 C
1701 732 C
1962 735 C
1962 734 C
2076 731 C
stroke
grestore
end
showpage
}
\put(1793,997){\makebox(0,0)[r]{$J_{m,0}^2\_dual$}}
\put(1793,1097){\makebox(0,0)[r]{$J_{m,0}^2$}}
\put(1793,1197){\makebox(0,0)[r]{$J_{m,i}^2\_dual$}}
\put(1793,1297){\makebox(0,0)[r]{$J_{m,i}^2$}}
\put(1318,51){\makebox(0,0){R [a]}}
\put(2156,151){\makebox(0,0){4}}
\put(1737,151){\makebox(0,0){3}}
\put(1318,151){\makebox(0,0){2}}
\put(899,151){\makebox(0,0){1}}
\put(480,151){\makebox(0,0){0}}
\put(420,1460){\makebox(0,0)[r]{1.5}}
\put(420,1218){\makebox(0,0)[r]{1}}
\put(420,976){\makebox(0,0)[r]{0.5}}
\put(420,735){\makebox(0,0)[r]{0}}
\put(420,493){\makebox(0,0)[r]{-0.5}}
\put(420,251){\makebox(0,0)[r]{-1}}
\end{picture}}
\vspace{-0.8cm}\caption{\label{j3}Comparison of $U(1)$ (at $\beta=0.96$) and the $\gamma \to \infty$ limit of the dual Higgs model: Transverse profile of the monopole current (left figure) and the squared spatial ($J_{m,i}^2$) and temporal ($J_{m,0}^2$) monopole currents (right figure) in the symmetry plane for charge distance $d=3a$. The error bars for the results obtained in the dual theory have been omitted because they are much smaller than the symbols.}
\end{figure*}
\begin{figure}
\vspace{-0.1cm}
\centerline{\input{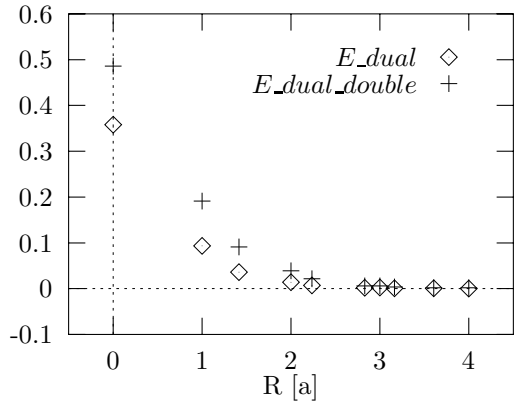}}
\vspace{-0.8cm}
\caption{\label{prof96}Electric field profile in dual $U(1)$ at $\beta=0.96$ between single and double charges in the symmetry plane for $d=3a$.}
\end{figure}
\begin{figure}
\vspace{-0.1cm}
\centerline{\input{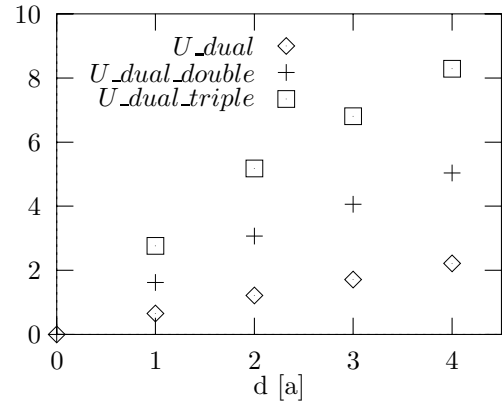}}
\vspace{-0.8cm}
\caption{\label{u96}Total energy as a function of charge distance for single, double and triple charges.}
\end{figure}

$U(1)$ lattice gauge theory in four dimensions undergoes a phase transition at strong coupling and therefore has been widely used to investigate the confinement mechanism. In $U(1)$ theory the condensation of magnetic monopoles is responsible for the confinement of electric charges. The results of lattice simulations can be interpreted in terms of the dual superconductor picture of confinement, e.g. the validity of a dual London relation has been checked\cite{haymaker,zach}.\\
On the other hand it was realized many years ago \cite{banks} that one can perform a duality transformation of the path integral. In this way a new partition function is obtained which can be regarded as a certain limit of a dual non-compact Abelian Higgs model \cite{froehlich}. We will extend the duality relation to expectation values in the presence of external charges, which should help to clarify the connection between dual transformed $U(1)$ theory and the dual superconductor model. \\
Our aim is to compare simulations in  the dual theory and in ordinary $U(1)$ gauge theory. In order to calculate expectation values in the original gauge theory with sufficient accuracy we use a very modest lattice size of $8^3 \times 4$. The results discussed below demonstrate that in the confinement phase the simulation of the dual theory is much more efficient, and calculations on larger lattices can be performed easily.\\
Expectation values of physical observables ${\cal O}$ in the presence of a static charge pair at a distance $d$ are determined by the correlation function
\begin{equation} \label{correlation}
\langle {\cal O}(x) \rangle_{Q\bar{Q}} = \frac{\langle L(0) L^\ast(d) {\cal O}(x) \rangle}{\langle L(0) L^\ast(d) \rangle} - \langle {\cal O} \rangle,
\end{equation}
where $L(\vec{r})$ is the Polyakov loop and the angle brackets denote the evaluation of the path integral using the ordinary Wilson action. The quantities we are interested in are the electric field strength and the monopole currents as well as their squares. For the field strength we use the identification 
$a^2 e F_{\mu\nu} = \sin \theta_{\mu\nu}$
which can be shown to fulfil the Gauss law for electric charges \cite{zach}. Then the monopole currents can simply be constructed via the dual Maxwell equations.\\
The results verify the dual superconductor scenario: The electric flux between a charge pair is squeezed into a flux tube, encircled by monopole currents acting like a coil. For the squared monopole currents, calculations on a Euclidean lattice show that both spatial and temporal components are suppressed in the flux tube. The sign of the spatial components, however, has to be changed when going back into Minkowski space \cite{zach}. A profile of monopole currents and squares of currents at the coupling $\beta=0.96$ is depicted in Fig.~\ref{j3}, to be compared to the dual theory.\\
How can the obtained results be understood in terms of the dual superconductor picture? A very general starting point is the non-compact Abelian Higgs model. Electric charges can be introduced by means of dual Dirac strings. The action in the limit of very large Higgs potential (extreme type II superconductor) reads
\begin{eqnarray} \label{higgs}
S &=& \beta \sum_{x,\;\mu > \nu} G\left({\rm d}\theta_{x,\,\mu\nu} + 2 \pi n_{x,\,\mu\nu}\right) - \nonumber \\
&-& \gamma \; \sum_{x,\;\mu} \; \frac{1}{2} \; \left( \Phi^*_x e^{- i \theta_{x,\,\mu}} \Phi_{x + \mu} + h.c.\right),
\end{eqnarray}
where the function $G(x)$ determines the non-compact gauge field action ($G(x) \to x^2/2$ in the continuum limit). $n_{x,\,\mu\nu}$ is fixed and describes the dual Dirac string connecting the charge pair. The complex Higgs field is constrained to $\Phi_{x}=\exp(i \chi_{x})$. The ratio $\beta/\gamma$ has the meaning of the squared London penetration length $\lambda$ in this model. The Higgs current $J = \gamma/\sqrt{\beta} \; ({\rm d}\chi - \theta)$  is constructed in such a way that the fluxoid quantisation $F + \lambda^2 {\rm d}J = m\,e$ is valid for every single field configuration. The integer two-form $m$ characterizes the physical fluxoid string which is allowed to fluctuate in simulations due to the compactness of the phase of the Higgs field.\\
Let us now briefly review how to connect this model to the dual transformed path integral of $U(1)$ gauge theory including external charges. Starting with the Wilson action and performing a Fourier expansion of the Boltzmann factor \cite{banks} one obtains the electric Gauss law as a constraint which is fulfilled by introducing integer link variables $^*p$ and plaquette variables $^*n$ on the dual lattice. One arrives at the dual partition function
\begin{equation}\label{zdual}
Z = \;\; (2 \pi)^{4N}  \prod \sum_{^*p}
\;\; e^{-\beta} I_{\parallel {\rm d} ^*p + ^*n \parallel}(\beta).
\end{equation}
This corresponds to the $\gamma \to \infty$ limit of the Higgs model (\ref{higgs}) written in dual variables, with $G(x)$ determined by the modified Boltzmann factor in eq.~(\ref{zdual}) and $\beta_{dual}=1/(4\pi^2\beta_{U(1)})$. The external electric current becomes the boundary of a dual Dirac sheet $^*n$ in the dual theory.\\
Finally we also perform the duality transformation on the correlation function (\ref{correlation}) to evaluate the expectation value of the electric field strength. It agrees with the electric field in the limit of the dual Higgs model:
\begin{equation}\label{ident}
< \vec{E} >_{Q\bar{Q}}^{U(1)} \; = \; < \vec{E} >_{Q\bar{Q}}^{dual \, Higgs \, (\gamma \to \infty)}.
\end{equation}
Due to the validity of Maxwell equations also the expectation values of the monopole currents agree. A comparison of numerical results can be seen in Fig.~\ref{j3} for charge distance $3a$. However, an identity like eq.~(\ref{ident}) does not hold for expectation values of squared fields and squared monopole currents. The right plot in Fig.~\ref{j3} shows that the two models behave similar, but there is no quantitative agreement.\\
The correspondence between $U(1)$ gauge theory and dual Higgs model in the limit $\gamma \to \infty$ (this means zero penetration length) allows for the interpretation that the finite extension of the flux tube is the result of fluctuations of Nambu's string. Due to the disagreement between squares, this picture might fail when looking at observables like the energy density.\\
Nevertheless, simulating the dual $U(1)$ gauge theory gives us the possibilty to calculate any expectation value with significantly higher precision. This has two reasons: Firstly, the confinement phase is the weakly coupled phase in the dual theory. Secondly, it is not necessary to project the charge--anticharge state out of the vacuum: Charge pairs with arbitrary distance can be simulated with equal accuracy, as well as multiple charges. As an example let us consider a pair of double charges at distance $d=3a$. The electric field profile in comparison to that of a singly charged pair is plotted in Fig.~\ref{prof96}. It can be seen that the flux tube widens for double charges. This is different from the results in deconfinement, where the field strength is simply doubled everywhere.
\\
Finally we calculate the total energy of the electromagnetic field as a function of charge distance for the Wilson action in $U(1)$ using the dual formulation. As mentioned above, this must not be confused with the energy of the dual Higgs model. A first calculation of this type was already performed in ref.~\cite{greensite} using the polymer representation of the path integral in three-dimensional $U(1)$. Whereas in the deconfinement phase one gets the expected Coulomb potential proportional to the square of the charges, the slope in the confinement phase scales proportionally to the charge (see Fig.~\ref{u96}). These results differ from those of ref.~\cite{trottier} which were obtained in three-dimensional $U(1)$ gauge theory.
\\
We conclude that the dual transformed theory both helps to interpret the behaviour of expectation values and can be of great use for precise calculations in the confinement phase of $U(1)$ lattice gauge theory. Further calculations in the dual theory are in progress and will be published elsewhere.


\begin{thebibliography}{9}
\bibitem{haymaker} V.~Singh, R.~W.~Haymaker, D.~A.~Browne, \pr\ {D47} (1993) 1715.
\bibitem{zach} M.~Zach, M.~Faber, W.~Kainz, P.~Skala, \pl\ B358 (1995) 325.
\bibitem{banks} T.~Banks, R.~Myerson, J.~Kogut, \np\ B129 (1977) 493.
\bibitem{froehlich} J.~Fr\"ohlich, P.~A.~Marchetti, Europhys.\ Lett. 2 (1986) 933.
\bibitem{greensite} T.~Sterling, J.~Greensite, \np\ B220 (1983) 327.
\bibitem{trottier} H.~D.~Trottier, R.~M.~Woloshyn, \pr\ D48 (1993) 2290.
\end{thebibliography}
\end{document}